\documentclass[12pt]{iopart}
\usepackage{amssymb, amsfonts, amsbsy}
\begin{document}
\def\TODAY{4 August 2010}
\title 
{Bi-metric pseudo--Finslerian spacetimes}
\author{
Jozef Skakala and 
Matt Visser}
\address{School of Mathematics, Statistics, and Operations Research, \\
Victoria University of Wellington, PO Box 600, Wellington, New Zealand}
\ead{jozef.skakala@msor.vuw.ac.nz, 
matt.visser@msor.vuw.ac.nz}
\date{\TODAY }
\begin{abstract}

Finsler spacetimes have become increasingly popular within the theoretical physics community over the last two decades. Because physicists need to use \emph{pseudo}--Finsler structures to describe \emph{propagation} of signals, there will be nonzero \emph{null vectors} in both the tangent and cotangent spaces --- this causes significant problems in that many of the mathematical results normally obtained for ``usual'' (Euclidean signature) Finsler structures either do not apply, or require significant modifications to their formulation and/or proof.  We shall first provide a few basic definitions, explicitly demonstrating the interpretation of bi-metric theories in terms of pseudo--Finsler \emph{norms}. We shall then discuss the tricky issues that arise when trying to construct an appropriate  pseudo-Finsler \emph{metric} appropriate to bi-metric spacetimes.  Whereas in Euclidian signature the construction of the Finsler metric typically fails at the zero vector, in Lorentzian signature the Finsler metric is typically ill-defined on the entire null cone.

\vskip 0.250cm


\vskip 0.1250cm

\bigskip
\noindent 
\TODAY;  
\LaTeX-ed  \today

\vskip 1cm
\hrule  
\tableofcontents
\bigskip
\hrule
\end{abstract}
\maketitle
\newtheorem{theorem}{Theorem}
\newtheorem{corollary}{Corollary}
\newtheorem{lemma}{Lemma}
\def\d{{\mathrm{d}}}
\def\implies{\Rightarrow}
\newcommand{\scri}{\mathscr{I}}
\newcommand{\sun}{\ensuremath{\odot}}
\def\ep{\epsilon}
\def\k{\mathbf{k}}
\def\x{\mathbf{x}}
\def\v{\mathbf{v}}
\def\s{\mathbf{s}}
\def\e{\mathbf{e}}
\def\t{\mathbf{t}}
\def\n{\mathbf{n}}
\def\u{\mathbf{u}}
\def\w{\mathbf{w}}
\def\eg{{\it e.g.}}
\def\ie{{\it i.e.}}
\def\etc{{\it etc.}}
\def\sign{{\hbox{sign}}}
\def\eof{\Box}
\newenvironment{warning}{{\noindent\bf Warning: }}{\hfill $\eof$\break}


\markboth{Bi-metric pseudo--Finslerian spacetimes}{}

\section{Introduction}

Over the last two decades, Finsler norms and Finsler metrics have become increasingly utilized in various extensions of general relativity, and sometimes in reinterpretations of more standard situations. However the fact that physicists need to work in Lorentzian signature $(-+++)$ instead of the Euclidean signature $(++++)$ more typically used by the mathematicians leads to many technical subtleties (and can sometimes completely invalidate naive conclusions).
After very briefly presenting the basic definitions, we shall  interpret bi-metric theories in a Finslerian manner, this being one of the simplest nontrivial Finsler structures one could consider.  While there is a very natural way of merging the two signal cones into  a ``combined'' pseudo-Finsler \emph{norm}, we shall see that the situation with regard to Finsler \emph{metrics} is considerably more complicated. 
To set the stage, we point out that in Bernhard Riemann's 1854 inaugural lecture~\cite{Riemann}, he made some brief speculations about possible extensions of what is now known as Riemannian geometry:
\begin{quote}
  {The next case in simplicity includes those manifolds in
    which the line-element may be expressed as the fourth root of a
    quartic differential expression.  The investigation of this more
    general kind would require no really different principles, but
    would take considerable time and throw little new light on the
    theory of space, especially as the results cannot be geometrically
    expressed\dots}
\end{quote}

\begin{quote}
  {\dots A method entirely similar may for this purpose be applied
    also to the manifolds in which the line-element has a less
    simple expression, \emph{e.g.}, the fourth root of a quartic
    differential.  In this case the line-element, generally speaking,
    is no longer reducible to the form of the square root of a sum of
    squares, and therefore the deviation from flatness in the squared
    line-element is an infinitesimal of the second order, while in
    those manifolds it was of the fourth order.}
\end{quote}
In more modern language, Riemann was speculating about distances being defined by  expressions of the form
\begin{equation}
\d s^4 = g_{abcd}  \; \d x^a \; \d x^b  \;\d x^c  \;\d x^d. 
\end{equation}
That is
\begin{equation}
\d s = \sqrt[4]{g_{abcd}  \; \d x^a \; \d x^b  \;\d x^c  \;\d x^d}. 
\end{equation}
Such manifolds, and their generalizations, have now come to be called
Finsler geometries~\cite{Finsler}. (More specifically, this particular case corresponds to a so-called 4th-root Finsler geometry.) 
Finsler geometries are extremely well-known in the mathematics community, with key references being~\cite{Finsler, Buseman, Rund, Chern1, Chern2, Chern3, Chern4, Shen1, Shen2, Antonelli93, Antonelli98, Bejancu}, but are considerably less common within the physics community~\cite{Gibbons, Kouretsis, Siparov, chang-li,  Gibbons2, dubrovnik, Girelli, Sindoni, Perlick,  Takano, Asanov, Brandt, Beil1, Bogoslovsky, Gonner, Vargas, Beil2, Mignemi}. Perhaps the most extensive use of pseudo-Finsler geometries has been within the ``analogue spacetime'' community~\cite{LRR} where Finsler-like structures have arisen in the context of normal mode analyses~\cite{normal1, normal2, inquiring}, and in multi-component BEC acoustics~\cite{lnp, trieste, silke-phd, bled}. 

\section{Finsler basics}

Mathematically, a Finsler function (Finsler norm, Finsler distance function) is defined as a function $F(x,v)$ on the tangent bundle to a manifold such that
\begin{equation}
F(x, \kappa \, v) = \kappa \; F(x, v).
\end{equation}
This then allows one to define a notion of distance on the manifold, in the sense that
\begin{equation}
S\left(x(t_i),x(t_f)\right) = \int_{t_i}^{t_f}  F\left( x(t),  {\d x(t) \over\d t} \right) \; \d t 
\end{equation}
is now guaranteed to be independent of the specific parameterization $t$. In the particular case of a (pseudo--)Riemannian manifold with metric $g_{ab}(x)$ one would take
\begin{equation}
F(x, v) = \sqrt{ g_{ab}(x)\; v^a\,v^b},
\end{equation}
but a general (pseudo--)Finslerian manifold the function $F(x, v)$ is completely arbitrary except for the linearity constraint in $v$. In Euclidean signature, the function $F(x,v)$ is taken to be smooth except at $v=0$. (This is most typically phrased in terms of the function $F(x,v)$ being smooth on the ``slit tangent bundle''; the tangent bundle with the zero vector deleted.) In Lorentzian signature however, we shall soon see that $F(x,v)$ is typically non-smooth for all null vectors --- so that non-smoothness issues have typically grown to affect the entire null cone. Sometimes a suitable \emph{power}, $F^n$,  of the Finsler norm is smooth. It is standard to define the (pseudo--)Finsler \emph{metric} as
\begin{equation}
g_{ab}(x,v) = {1\over2} \; { \partial^2 [F^2(x,v)] \over \partial v^a \; \partial v^b}
\end{equation}
which then satisfies the constraint
\begin{equation}
g_{ab}(x, \kappa\, v ) = g_{ab}(x, v).
\end{equation}
This can be viewed as a ``direction dependent metric'', and is clearly a generalization of the usual (pseudo--)Riemannian case. Almost all of the relevant mathematical literature has been developed for the Euclidean signature case (where $g_{ab}(x,v)$ is taken to be a positive definite matrix).  Herein we wish to raise some cautionary flags with regard to the Lorentzian signature pseudo-Finsler case. 

\section{Bimetric theories}

Bi-metric theories contain two distinct metrics $g^\pm_{ab}$, so we can define two distinct ``elementary'' Finsler norms $F_\pm(x,v) = \sqrt{g^\pm_{ab} \; v^a v^b}$.  
Suppose one now wants a combined Finsler norm that simultaneously encodes both signal cones --- then the natural thing to do is to implement Bernhard Riemann's original suggestion and take
\begin{equation}
F(x,v) = \sqrt{ F_+(v) \; F_-(v)}; \qquad g_{abcd} = g^+_{(ab} \; g^-_{cd)}.
\end{equation}
This construction for $F(x,v)$ is automatically linear in $v$, and the vanishing of $F(x,v)$ correctly encodes the two signal cones. So this definition of $F(x,v)$ provides a perfectly good Finsler \emph{norm}. The Finsler \emph{metric} is however quite ill behaved, and the technical problems can be traced back to the fact that we are working in Lorentzian signature --- the issues we discuss below have no real impact in Euclidean signature.

First, note that the norm $F(x,v)$ has the interesting ``feature" that it picks up non-trivial complex phases: Since $F_\pm(x,v)^2$ is always real,  (positive inside the propagation cone, negative outside),  $F_\pm(x,v)$ is either pure real or pure imaginary. But then, thanks to the additional square root in defining  $F(x,v)$, one has:
\begin{itemize}
\item  $F(x,v)$  is pure real inside both propagation cones.
\item  $F(x,v)$  is proportional to $\sqrt{i} = {1+i\over\sqrt2}$  between the two propagation cones.
\item  $F(x,v)$  is pure imaginary outside both propagation cones.
\end{itemize}
Thus in a bi-metric Lorentzian signature situation the particular ``natural'' pseudo-Finsler norm considered above \emph{cannot} be smooth as one crosses the propagation cones --- what was in Euclidean signature a complicating technical feature that only arose at the zero vector of each tangent space has in Lorentzian signature grown to affect all null vectors. This pseudo-Finsler norm is at best ``smooth on the tangent bundle excluding the null cones''. (Note that individually the $F_\pm(x,v)^2$ are smooth across the propagation cones, but in the bi-metric case one has to go to $F(x,v)^4$ to get a smooth function.)
By extension, it is clear that similar phenomena will occur whenever one encounters multi-sheeted signal cones (bi-refringence, multi-metric spacetimes, multi-refringence).

Second, when attempting to bootstrap this Finsler \emph{norm} to a pseudo-Finsler \emph{metric} one encounters additional and more significant complications. 
The Finsler metric
will have (at least some) infinite components --- $g_{ab}(x,v)$ has infinities on both signal cones. 
To see this consider
\begin{eqnarray}
\fl
g_{ab}(x,v) &=& {1\over2} \; \partial_a \partial_b \sqrt{ F_+^2 \; F_-^2 }
\\
\fl
&=& {1\over4} \partial_a \left[ \partial_b [F_+^2] \; {F_-\over F_+} 
+  \partial_b [F_-^2]\;  {F_+\over F_-} \right]
\\
\fl
&=& {1\over4}  \left[ \partial_a \partial_b [F_+^2] \; {F_-\over F_+} + 
 \partial_b \partial_b [F_-^2]  \; {F_+\over  F_-} \right]
\nonumber
\\
\fl
&&+ {1\over2} \left[  
{\partial_a F_+ \partial_b F_- + \partial_a F_- \partial_b F_+ }
-
\partial_a F_+ \partial_b  F_+ \;{F_-\over F_+} 
-
\partial_a F_- \partial_b F_- \;{F_+\over F_-}
\right]\!\!.
\end{eqnarray}
That is, tidying up:
\begin{eqnarray}
\fl 
g_{ab}(x,v) &=& {1\over2} \left[ g^+_{ab}\; {F_2\over F_1} 
+  g^-_{ab}\; {F_1\over F_2} \right]
\nonumber\\
\fl &&
+ {1\over2} \left[  {\partial_a F_+ \partial_b F_-+ \partial_a F_- \partial_b F_+}
-
\partial_a F_+ \partial_b F_+ \; {F_-\over F_+} 
-
\partial_a F_- \partial_b F_- \; {F_+\over F_-}
\right]\!\!.
\end{eqnarray}
The problem is that this ``unified''  and ``natural'' Finsler metric $g_{ab}(x,v)$ has singularities on both of the signal cones.

The good news is that the quantity $g_{ab}(x,v) \, v^a v^b = F^2(x,v)$, and so since on either propagation cone $F(x,v)\to 0$, we see that $F(x,v)$ itself has a well defined limit. 
But now let $v^a$ be the vector the Finsler metric depends on, and let  $w^a$ be some \emph{other} (arbitrary) vector.
Then
\begin{eqnarray}
\fl g_{ab}(x,v)  \, v^a w^b &=&  {1\over2} v^a w^b \partial_a \partial_b [F^2] = {1\over2} w^b \partial_b [F^2] = {1\over2} w^b \partial_b \sqrt{F_+ F_-}
\nonumber
\\
\fl
&=& {1\over4} w^b \left[ \partial_b [F_+^2] \; {F_-\over F_+} +  
\partial_b [F_-]^2 \; {F_+\over F_-} \right]
\nonumber
\\
\fl
&=& {1\over2} \left\{ (\,g^+_{ab} \, v^a w^b) \; {F_-\over F_+} +
 (\,g^-_{ab} \, v^a w^b) \; {F_+\over F_-} \right\}.
\end{eqnarray}
The problem now is this: $g^+_{ab}$ and $g^-_{ab}$ are both by hypothesis individually well defined and finite. But now as we go to propagation cone ``$+$'' we have
\begin{equation}
g_{ab}(x,v) \, v^a w^b \to {1\over2}  (\,g^+_{ab} \; v^a w^b) \;  {F_-\over 0}  = \infty,
\end{equation}
and as we go to the other propagation cone ``$-$'' we have
\begin{equation}
g_{ab}(x,v) \, v^a w^b \to {1\over2}  (\,g^-_{ab} \; v^a w^b) \;  {F_+\over 0}  = \infty.
\end{equation}
So at least some components of this ``unified'' Finsler metric $g_{ab}(x,v)$ are unavoidably singular on the propagation cones. Related singular phenomena have been encountered in multi-component BECs, where multiple phonon modes can interact to produce Finslerian propagation cones~\cite{lnp}. 
Things are just as bad if we pick $u$ and $w$ to be \emph{two} vectors distinct from $v$. Then
\begin{eqnarray}
\fl
g_{ab}(x,v) \, u^a w^b &=& {1\over2} \Bigg[ 
g^+(u,w) \;{F_-\over F_+} + g^-(u,w) \;{F_+\over F_-} 
\nonumber
\\
\fl
&&
+ {g^+(u,v) \, g^-(w,v) + g^+(w,v) \, g^-(u,v)\over F_+ \, F_-} 
\nonumber
\\
\fl
&&
- g^+(u,v) \, g^+(w,v) \; {F_-\over F_+^3} - g^-(u,v) \, g^-(w,v) \; {F_+\over F_-^3}
\Bigg].
\end{eqnarray}
Again, despite the fact that $g^+$ and $g^-$ are by hypothesis regular on the signal cones. the ``unified'' Finsler metric $g_{ab}(x,v)$ is unavoidably singular there --- unless, that is, you \emph{only} choose to look in the $vv$ direction. By extension, it is clear that similar phenomena will occur whenever one encounters multi-sheeted signal cones (bi-refringence, multi-metric spacetimes, multi-refringence).

\section{Discussion and Conclusions}

On the one hand we have seen how bi-metric theories are good exemplars for providing a clean physical implementation of  the mathematical notion of a pseudo-Finsler norm,  and how they naturally lead to a prescription for building a pseudo-Finsler metric. On the other hand we have also seen how this rather straightforward physical model nevertheless leads to significant technical mathematical difficulties.

It is when one tries to ``unify'' the two metrics into a single structure that 
the most significant  problems arise ---  the  pseudo-Finsler \emph{norm} is certainly well defined (and is extremely close to Riemann's original conception of what a 4th-order geometry should look like), but now the pseudo-Finsler \emph{metric}  is singular on the entire signal cone.  This problematic feature is intimately related to the fact that we are dealing with Lorentzian signature \emph{pseudo}-Finsler geometries --- it is a ``divide by zero'' problem, associated with non-zero null vectors on the signal cone, that leads to singular values for metric components. This appears to us to be an intrinsic and unavoidable feature of  bi-metric \emph{pseudo}-Finsler spacetimes. 

In earlier work~\cite{Jozef1, Jozef2} we had investigated this point within the technically more complicated context of biaxial birefringent crystal optics. The many extra technical details required to deal with biaxial crystals somewhat obscured the generality of the point we wish to make. The considerably simpler framework of bi-metric theories is sufficient to make our point with clarity: pseudo-Finsler metrics are typically not smooth over the entire null cone.

In closing we wish to emphasize that this does not mean that all attempts at constructing pseudo-Finsler metrics are intrinsically ill conceived. The situation is more subtle. While the situation with multiple signal cones is clearly diseased, at least if one wishes to encode all signal cones in one Finsler structure in any straightforward manner, and while one cannot blindly carry Euclidean signature Finsler results over to Lorentzian signature, the case of a single (geometrically distorted) signal cone may still be of interest --- one will just have to check explicitly that all Euclidean signature constructions can be generalized to Lorentzian signature.  While this step is relatively straightforward for the Riemannian $\to$ pseudo-Riemannian transition, it is much more subtle for the Finsler $\to$ pseudo-Finsler transition. 

\ack

This research was supported by the Marsden Fund administered by the
Royal Society of New Zealand. 
JS was also supported by a Victoria University of Wellington postgraduate scholarship.

\section*{References}
\addcontentsline{toc}{section}{References}


\end{document}